# Partial Fourier techniques in single-shot cross-term spatiotemporal encoded MRI

Zhiyong Zhang and Lucio Frydman*

*Department of Chemical Physics, Weizmann Institute of Science, Rehovot 76100, Israel*

*Prof. Lucio Frydman; +972-8-9344903; lucio.frydman@weizmann.ac.il

# ABSTRACT

**Purpose:** Cross-term spatiotemporal encoding **(**xSPEN) is a single-shot imaging approach with exceptional resilience to field heterogeneities: its images do not require *a priori* information nor use post-acquisition corrections, to deliver faithfully the spins' spatial distribution. xSPEN, however, suffers from SNR penalties due to its non-Fourier nature and due to diffusion losses – especially when desiring high resolution. This study explores partial Fourier transform approaches that acting along either the readout or the spatiotemporally-encoded dimensions, reduce these penalties.

**Methods:** xSPEN uses an orthogonal (e.g., *z*) gradient to read, in direct space, the low-bandwidth (e.g., *y*) dimension. This changes substantially the nature of partial Fourier acquisitions vis-à-vis conventional imaging counterparts. A suitable theoretical analysis, however, allows one to implement these procedures along either the low-bandwidth or readout axes.

**Results:** Partial Fourier single-shot xSPEN methods are illustrated on preclinical and human scanners. Owing to their reduction in the experiments' acquisition times, these methods provide substantial sensitivity gains vis-à-vis previous implementations for a given targeted in-plane resolution. The magnitude of these gains is as expected.

**Conclusion:** Partial Fourier approaches, particularly when implemented along the xSPEN ("low-bandwidth") axis, can provide substantial sensitivity advantages at minimal costs in the operation of the single-shot experiments.

**INTRODUCTION**

Cross-term spatiotemporal encoding (xSPEN) is a recently introduced approach delivering single-scan NMR images with unprecedented resilience to field inhomogeneities (1). Like its spatiotemporally-encoded (SPEN) predecessors (2-5), xSPEN relies on imprinting a shaped phase during an initial encoding process, which then serves as focal point for a subsequent, gradient-driven image readout. Both of these methods differ from $k$-based scans like Echo Planar Imaging (EPI) (6,7), in that their image readout happens in direct, physical space. To do so a strong non-linear phase profile $\varphi(r)$ is imparted that leads to destructive interferences among signals emitted from neighboring spins –except for those positioned close to positions fulfilling the stationary-phase condition $(\nabla \varphi)_{r=r_o} = 0$. $(\partial \varphi / \partial r)_{r=r_o} = 0$. The action of an acquisition gradient $G_a$ will provide then an additional evolution $\varphi_a = k_a \cdot r$ with $k_a = \gamma G_a t$, that displaces this stationary phase point throughout the targeted Field of View (FOV). If properly steered, this will eventually reveal the full $\rho(r)$ spin density over the targeted FOV over the course of the acquisition. SPEN imparts such encoding in the form of a quadratic, $y^2$-type phase modulation, whereas xSPEN does it by relying on a hyperbolic $y \cdot z$-type phase (1,8). As a result of the latter the option arises of activating either $G_a^y$ or $G_a^z$ acquisition gradients that will unravel, respectively, either the $\rho(z)$ or the $\rho(y)$ spatial profiles. The physical basis of how the application of a given acquisition gradient allows one to read, in direct space, the spins' profile along an orthogonal axis, has been explained in detail elsewhere (1). As also detailed in Ref. (1) the second of these options, which utilizes a $z$-axis $G_a^z$ gradient to both encode and also to unravel a $\rho(y)$ image, enables one to perform a peculiar image acquisition that can be entirely free from chemical shift or field inhomogeneity effects. This reflects the fact that, rather than viewing frequency dispersions as artifacts that need to be overcome by the application of an overwhelming external field gradient, this way of performing MRI incorporates any disturbing frequency broadening as part of both the initial hyperbolic phase encoding and the subsequent image decoding processes. This capability is particularly valuable when considering single-shot 2D acquisitions, experiments that although central in numerous diffusion- and functional-oriented applications, are known to be particularly sensitive to field inhomogeneity distortions (9-11). Figure 1a illustrates one of the ways whereby the xSPEN strategy just described, can be adapted for the realization of such single-shot 2D acquisitions. To impart its hyperbolic phase encoding this sequence turns on a $G_z$ along the slice-selection axis; this is used for the initial excitation of a slice with width $L_z$, and is kept on throughout the rest of the scan. In combination with two linearly-swept adiabatic inversion pulses (4,12) applied in the presence of a bipolar gradient $\pm G_y$, this results in the $\varphi_e = -Cyz$ phase profile that characterizes xSPEN's encoding –where $C$ is a spatiotemporal encoding constant under the experimentalist's control, and $y, z$ are positions in the $-FOV_y/2 \leq y \leq FOV_y/2$, $-L_z/2 \leq z \leq L_z/2$ ranges. Then, over the course of the acquisition, the continued action of the constant $G_z$ displaces this saddle-shaped profile along the $y$-axis progressively probing $\rho(y)$; in synchrony with this, an oscillating $\pm G_x$ gradient applied along an orthogonal readout dimension explores the $k_x$-axis in a conventional, EPI-like manner. The mechanism by which the constant application of a $G_z$ gradient delivers an image free from inhomogeneities has been discussed in detail elsewhere (1,13); basically, even if an inhomogeneity $\delta\omega(r)$ distorts the encoding phase from $\varphi_e = -Cyz$ into $\varphi_e = -Cy(z+\delta\omega/\gamma G_z)$, the subsequent decoding will take place aided by the same inhomogeneity. This lead to a signal $S$ that as a function of the oscillating wavenumber $k_x$ and of the acquisition time $t$, can be expressed as

$$S(k_x,t) = \int_X dx \cdot e^{ik_x x} \int_Y dy \cdot \rho(x,y) \cdot \frac{L_z}{1+f[\delta\omega]} \cdot \text{sinc}\left[(-Cy+\gamma G_z t)\frac{L_z}{2}\right]. \qquad (1)$$

This equation means that rearrangement of the resulting data and 1D Fourier transformation (FT) along $k_x$ will lead, apart from a potential distortion related to the slice-selection and represented by the $f$ function, a 2D $\rho(x,y)$ image as a function of $t$ that will be devoid from all inhomogeneity effects.

While capable of delivering single shot images devoid from in-plane distortions, xSPEN's lack of FT along the low bandwidth dimension imparts a substantial signal-to-noise ratio (SNR) loss (1). This SNR penalty is compounded by the constant $G_z$ gradient required by xSPEN, which being larger than a usual EPI phase-encoding gradient by a ratio $\approx \frac{FOV_y}{L_z}$, will impart additional losses. All this makes single-shot xSPEN considerably less sensitive than methods like EPI or even its SPEN predecessors –particularly if operating under moderate inhomogeneities and for the long acquisition times $T_a$ required for high in-plane resolutions. A well-known approach to alleviating such effects relies on the partial FT (pFT) (14,15); an approach that leverages the properties of the image being sought, in order to reduce the acquisition coverage along one of the $k$-domains. Indeed it is a basic feature of NMR that given that given the real-nature of the spectral correlations, it is in principle possible to sample only half the extent of the full *k*-space, and still achieve the same levels of spatial resolution as would arise from sampling a full $-k^{max} \leq k \leq k^{max}$ range of values (16,17). In practice such maximal reduction in the sampled data is rarely achieved, and partial sampling factors $0.6 \leq p \leq 0.8$ are more common. The $T_a \rightarrow p \cdot T_a$ shortening of the overall acquisition times associated to this partial sampling can lead to a considerable reduction in relaxation and in diffusion-driven losses – particularly for long-$T_a$, constant-gradient sequences such as xSPEN. The question then arises of how to exploit these *k*-based phase-conjugation arguments in sequences that, like SPEN or xSPEN, are based on the hybrid sampling of $k_x$ and of *y*-domains. pFT considerations are in fact directly applicable to the first of these domains, and the ensuing $T_a$ reduction associated to a reduced sampling of the readout axis leads to the expected SNR gains in the xSPEN scan. Less evident but in fact even more beneficial are the options arising upon adopting partial Fourier methods to imaging the xSPEN axis; the physical basis of both experiments and demonstrations of their ensuing usefulness to achieve resolutions that so far had been out of reach for single-shot xSPEN experiments in realistic conditions, are presented below.

**METHODS**

*Theoretical background.* As mentioned, pFT seeks to retain spatial resolution while reducing MRI's acquisition times, by estimating part of the *k*-space data based on complex conjugation. Thus, although the inherent resolution of a 1D MRI acquisition depends on the maximal sampled wavenumber |$k^{max}$|, phase distortions and blurring will characterize magnitude images unless a symmetric $-k^{max} \leq k \leq +k^{max}$ region is sampled. Partial FT procedures rely on the fact that for real spatial distributions the corresponding *k*-domain data have to fulfill $S(-k) = [S(+k)]^*$ (14,18), in order to calculate the image that would arise from the full $-k^{max} \leq k \leq +k^{max}$ support while limiting actual samples to a $0 \approx k \leq +k^{max}$ fraction. When extending these considerations from a 1D axis to a 2D plane, two potential strategies emerge. One is to exploit the $S(-k_x,-k_y) = [S(k_x,k_y)]^*$ symmetry along the directly-detected readout domain; the other is to apply it along the phase-encoded dimension. In conventional multi-shot MRI the latter is the preferred option, as it may shorten by the full *p*-factor the duration of the experiment. Single-shot techniques like EPI also generally apply pFT only along the phase-encoded domain, as doing so along the readout axis tends to complicate even/odd artifact corrections. In single-shot xSPEN imaging the readout dimension (*x*) is *k*-based, while the low bandwidth dimension (*y*) is space-based. Thus a pFT$_x$ can be carried out long each of the partially sampled readout axes, and all

data –including the two sets collected as a function of $-k_x$ and $+k_x$ gradients– combined in a $\rho(x,y)$ image space without additional complications. This is to some extent simpler than what is generally needed when implementing a $pFT_x$ in EPI-based sequences. Here, readout-based pFT procedures make the correction of even/odd artifacts even more challenging than in full-echo EPI; by contrast, xSPEN's low-bandwidth dimension is spatiotemporal decoded in direct spatial space, with no need for FT processing and thus devoid from even/odd corrections. In xSPEN, a $pFT_x$ readout reconstruction is done separately on positive and negative $k_x$-axis acquisitions, and the two datasets combined in image space without suffering from phase problems (Fig. 1b).

Less straightforward is envisioning how pFT could be exploited along the xSPEN $y$-dimension. As mentioned single-shot xSPEN imparts a preacquisition hyperbolic phase-encoding $e^{-iCyz}$, whose stationary point is shifted over the course of the acquisition by a constant $z$-gradient. Such gradient performs in essence an "analog Fourier analysis" on the encoded data while simultaneously removing all $\Delta B_o$ inhomogeneity effects, thereby delivering a $y$-axis image. This in turn means that an inverse FT of the data collected while under the action of xSPEN's $G_z$ gradient, will be the equivalent of a conventionally $k_y$-encoded MRI acquisition, with $k_y=-Cz$ being the Fourier-conjugate to the $y$-position. Therefore, in the same way as conventional pFT relies on breaking the echo symmetry of the $k$-domain acquisition by applying a pre-winding gradient, performing an "asymmetric" encoding of the xSPEN image would demand the introduction of a pre-winding $G_y$ pulse –even if the image is subsequently unraveled by the action of a $z$ acquisition gradient. Figure 1a highlights how this route to performing $pFT_y$ along the low-bandwidth dimension can be included into the original xSPEN 2D sequence, by introducing a short prephasing pulsed gradient $k_y^0$ along $y$. Such prephasing will effectively shift the "virtual $k_y$" encoding that originates the xSPEN signal. To see how this can lead to an enhanced image resolution for identical acquisition conditions, we revisit Eq. (1) in the absence of inhomogeneities and for a 1D case that ignores for the time being the readout dimension. Approximating the *sinc* function in this formula as

$$L_z \text{sinc}\left[(-Cy+\gamma G_z t)\frac{L_z}{2}\right] \approx \int_{-\frac{L_z}{2}}^{+\frac{L_z}{2}} dz \cdot e^{i(-Cy+\gamma G_z t)z} \qquad (2)$$

enables one to describe the effect of the prephasing pulsed gradient $k_y^0$ on the detected signal as

$$S[k_z(t)] = \int_Y dy \int_Z \rho(y) e^{-ik_y^0 y} e^{i(-Cy+k_z)z} \, dz \approx e^{-ik_y^0 y'} r(y'), \tag{3}$$

where $y' = k_z/C$ is the coordinate decoded by the action of the acquisition wavenumber $k_z = \gamma G_z t$, and $r(y')$ is a function representing the xSPEN-afforded image given by a convolution of the $\rho(y)$ spin density with the *sinc*-based sampling point spread function. The $e^{-ik_y^0 y'}$ prefactor multiplying this image clearly represents a shift in the $k_y$-space origin associated with $r(y')$'s inverse Fourier transform signal $s(k_y) = \int_Y r(y') e^{ik_y y'} dy'$. In other words, if in conventional xSPEN the maximum y-axis spatial resolution is given by the *sinc*'s width $\frac{2}{CL_z}$, the equivalent $k_y$ sampling associated to the prefactor in Eq. (2) will be shifted from $-CL_z/2 \leq k_y \leq CL_z/2$ to a $-CL_z/2 + k_y^0 \leq k_y \leq CL_z/2 + k_y^0$ interval. Hence, an inverse FT of the acquired xSPEN image, a suitable phase-conjugation processing, and a forward FT, should yield images with an extended $k_y$-support and hence an enhanced y-axis resolution.

These arguments hold for the case of a 1D xSPEN acquisition. Similar pFT considerations will apply to single-shot 2D experiments if the imaging processes along xSPEN and readout axes are fully decoupled; this will be the case if, for instance, the $G_z$ acquisition gradient would be pulsed in-between the bipolar readout gradients. In practice, however, it is often convenient to leave $G_z$ on continuously, as this frees not only the low-bandwidth but also the readout dimension from field inhomogeneity distortions. The simultaneous action associated with the oscillating $G_x$ and the constant $G_z$ gradients acting during xSPEN's 2D single-shot acquisition, however, brings about new features that need to be corrected before attempting a pFT$_y$. For introducing these features and deriving their pre-processing corrections, we consider for simplicity an xSPEN evolution that is free from relaxation, diffusion or field inhomogeneities. The 2D signal observed in the experiment can then be expressed as

$$S(k_x, k_z) = \begin{cases} S^{odd}(k_x, k_z) = \iiint_{X\,Y\,Z} \rho(x,y) e^{i\phi^{odd}} e^{-i(Cz\cdot y + k_y^0 y)} e^{ik_x x} e^{i(k_z z + \beta k_x z)} \, dx\,dy\,dz & \text{if } G_x \geq 0 \\ S^{even}(k_x, k_z) = \iiint_{X\,Y\,Z} \rho(x,y) e^{i\phi^{even}} e^{-i(Cz\cdot y + k_y^0 y)} e^{ik_x x} e^{i(k_z z - \beta k_x z)} \, dx\,dy\,dz & \text{if } G_x < 0. \end{cases} \tag{3}$$

Here the integrals extend over the FOVs targeted along the *x,y* axes as well as over the $L_z$ slice, $k_z$ and $k_x$ are the acquisition wavenumbers along the low-bandwidth and readout axes, $\beta$ is a "zigzag" factor (17,19) reflecting the fact that the $k_x$ wavenumber advances/recedes in conjunction with progress along $k_z$ over the course of the readout oscillation, and $\phi^{odd}$, $\phi^{even}$ are unknown phase terms associated to imperfections in the readout gradients. Adapting the $s(k_y) = \int_Y r(y')e^{ik_y y'} dy'$ notation introduced above to this 2D sampling case, we introduce functions associated to what would be the conventional $k$-space signal associated to this acquisition; i.e.

$$s^{odd}(k_x, k_y) = \int_X \int_Y \rho(x,y) e^{i\phi^{odd}} e^{ik_x x} e^{ik_y y} \, dx \, dy$$
$$s^{even}(k_x, k_y) = \int_X \int_Y \rho(x,y) e^{i\phi^{even}} e^{ik_x x} e^{ik_y y} \, dx \, dy. \qquad (4)$$

These "virtual signals" arising from positive and negative readout gradients can be used to rewrite Eq. (4) as

$$S(k_x, k_z) = \begin{cases} S^{odd}(k_x, k_z) = \int_Z s^{odd}\left(k_x, k_y + k_y^0\right) e^{i(k_z z + \beta k_x z)} dz \\ S^{even}(k_x, k_z) = \int_Z s^{even}\left(k_x, k_y + k_y^0\right) e^{i(k_z z - \beta k_x z)} dz \end{cases}. \qquad (5)$$

Furthermore, since $k_z$ rasterizes the $y$-axis, this is equivalent to the mixed-domain interferogram

$$S(k_x, y') = \begin{cases} S^{odd}(k_x, y') = \int_{k_y} s^{odd}\left(k_x, k_y + k_y^0\right) e^{i(k_y y' + \beta k_y k_x /C)} dk_y \\ S^{even}(k_x, y') = \int_{k_y} s^{even}\left(k_x, k_y + k_y^0\right) e^{i(k_y y' - \beta k_y k_x /C)} dk_y. \end{cases} \qquad (6)$$

where $y' = k_z/C$. If it would not be for the $\beta$-related terms, one could apply the same arguments that followed Eq. (3) in order to justify and achieve from these data a pFT$_y$-enhanced resolution. To better appreciate the effects associated to $\beta$'s "zigzag", we perform on Eq. (7) a final change of variables $k_y' = k_y + k_y^0$ and arrive to

$$S(k_x, y') = \begin{cases} S^{odd}(k_x, y') = e^{-ik_y^0 y'} e^{-i\beta k_x k_y^0/C} \int_{k_y'} s^{odd}\left(k_x, k_y'\right) e^{ik_y'(y' + \beta k_x/C)} dk_y' \\ S^{even}(k_x, y') = e^{-ik_y^0 y'} e^{i\beta k_x k_y^0/C} \int_{k_y'} s^{even}\left(k_x, k_y'\right) e^{ik_y'(y' - \beta k_x/C)} dk_y'. \end{cases} \qquad (7)$$

The $e^{-ik_y^0 y'}$ phase-modulation term here is as in conventional pFT$_y$; however, the new phase terms $e^{i\beta k_x k_y^0/C}$ and $e^{-i\beta k_x k_y^0/C}$ affecting the $S^{even}$ and $S^{odd}$ interferograms, evidence a coupling between the $k_y^0$ echo shifts and the $k_x$ sampling, that needs to be removed from even and odd data sets before performing a pFT$_y$ processing. In practice we apply this "zigzag correction" procedure, involving a row-by-row multiplication of *a priori* known phase terms, in conjunction with a removal of the $e^{i\phi^{odd}}$ and $e^{i\phi^{even}}$ phase imperfections that may affect signals collected under $\pm G_x$ gradients (17,20,21). (Fixing these even/odd phase problems is not essential in the original xSPEN experiment where solely a 1D FT along the readout axis is involved, yet hen implementing the $k_y^o$ encoding the additional manipulations involved in the pFT require accounting for them). The full procedure, involving the aforementioned zigzag+even/odd post-acquisition phase corrections followed by the pFT complex conjugation procedure, is summarized and exemplified in Figure 1c. In the present study the POCS (Projection Onto Convex Sets) partial Fourier reconstruction (22,23), was the pFT algorithm chosen to enhance resolution along either the readout or low-bandwidth axes.

*Experimental.* Phantom and animal-based acquisitions were carried out on a 7T/120mm horizontal magnet using a quadrature birdcage coil probe and DD2 Agilent Console (Agilent Technologies, Santa Clara, CA). Animal protocols and maintenance were done in accordance with guidelines of the Institutional Committee on Animals of the Weizmann Institute of Science (IACUC protocol #10790514). Spin-echo multi-shot (SEMS) imaging and SE-EPI experiments were carried out using sequences taken from the scanner's library; all SE-EPI acquisitions required reference "navigator" scans for affording reasonable images. SPEN and xSPEN imaging experiments were run in this preclinical scanner using custom-written pulse sequences and processing macros that were integrated into Agilent/Varian's VNMRJ® imaging software; these are available upon request. Human volunteers were also scanned on a 3T Siemens TIM TRIO platform using a 32-channels head coil. Compared in these scans were SE-EPI sequences taken from the scanner's library, against custom-written xSPEN acquisition/ processing programs. The experiments were approved by the Internal Review Board of the Wolfson Medical Center (Holon, Israel) and collected after obtaining informed, suitable-written consent. Main parameters used for setting up the various experiments are detailed in the corresponding figures' captions.

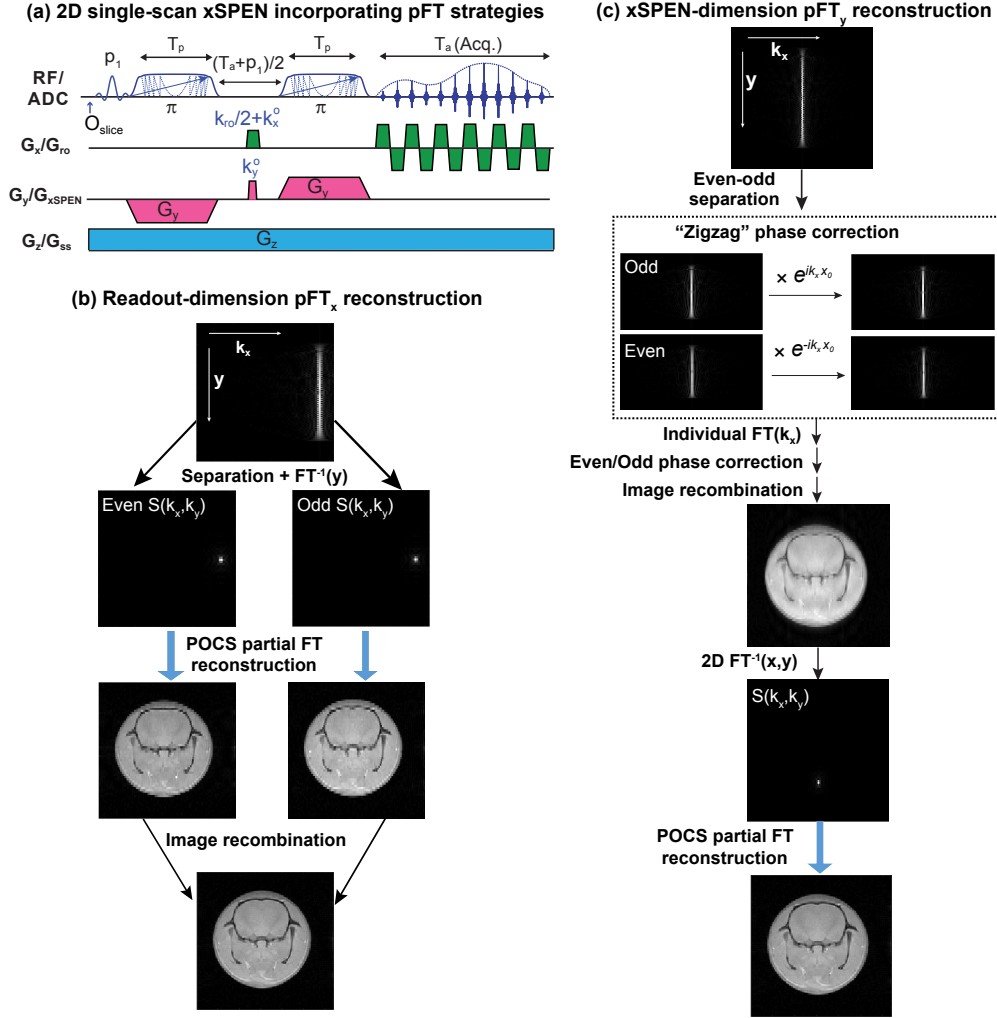

**Figure 1.** (a) Single-shot xSPEN sequence incorporating the possibility of implementing partial Fourier acquisitions by adding short prephasing pulsed gradients along the readout (pFT$_x$) or xSPEN (pFT$_y$) axes. (b) pFT$_x$ reconstruction involving the addition of a $k_x^0$ gradient pulse that displaces the S($k_x$,y) interferogram (top), separate processing of even/odd S($k_x$,$k_y$) data sets via POCS reconstruction, and subsequent combination (interleaving of magnitude data in image space to avoid phase problems) of the two sets. (c) pFT$_y$ reconstruction involving the addition of a $k_y^0$ gradient pulse that modulates the xSPEN y-image; separation of even/odd data sets; phase correction by *a priori* known zigzag effects $k_x x_o$ with $x_o = \beta k_y^o / C$; subsequent correction of residual even/odd phase problems; and final POCS-based partial FT reconstruction of the effective *k*-domain S($k_x$,$k_y$) data.

## RESULTS

Figure 2 illustrates the advantages that can result from adopting the pFT procedures just discussed, with a series of xSPEN experiments performed on a 7T preclinical scanner. Analyzed in these experiments was a lime, onto which a non-ferromagnetic titanium screw of a kind usually employed in orthopedic prostheses, had been inserted at a point of nearly axial symmetry for exacerbating the field inhomogeneities. Figure 2a shows a picture of the screw+fruit together with a SEMS sagittal image showing the effects of the titanium, as well as a challenging slice on which further axial analyses (Figs. 2b-2i) where implemented. These included a comparison between SEMS imaging (usually used as our "gold-standard", Fig. 2b) and images collected with SE-EPI, with fully-refocused SPEN and with the xSPEN sequence introduced in (1); this Fig. 2c-2e progression clearly shows the latter's higher robustness and faithfulness. Using this single-shot xSPEN image collected with the original sequence as starting point, Figures 2f-2i illustrate the kind of improvements that can be achieved by implementing pFT procedures. Figures 2f and 2g show images obtained upon breaking the symmetry of xSPEN's readout acquisitions, and reducing the number of points collected for each segment from 64 to 40. While a simple $FT_x$ procedure yields a lower resolution vis-à-vis the original 64-points xSPEN acquisition (Fig. 2e), the $pFT_x$ processing clearly restores this resolution. At the same time, the shortened echo times brought about by the 64→40 =0.625 reduction in readout points, brings a clear improvement in sensitivity. An even larger sensitivity improvement is observed, for identical reduction values, if the pFT procedure is implemented along the low-bandwidth dimension. Indeed, while Figure 2h shows once again that resolution is sacrificed upon reducing the sampled xSPEN lines from 64→40, Figure 2i demonstrates how the procedure introduced in Figure 1c can restore the lost resolution, while nearly tripling SNR vis-à-vis the original single-shot xSPEN sequence (Figs. 2i vs 2e). This is a general feature of these experiments: better SNR improvements can be achieved if applying the pFT along the xSPEN rather than along the readout dimension.

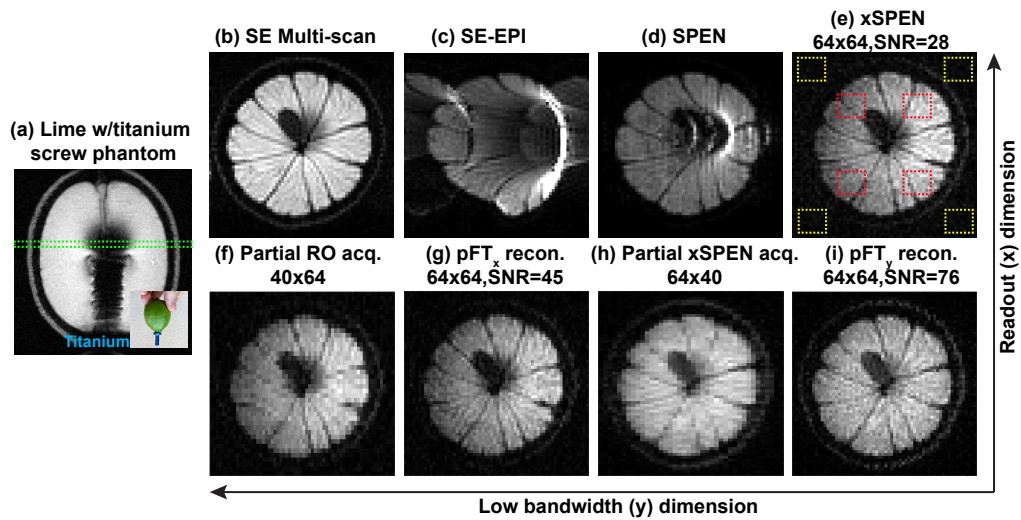

**Figure 2.** Representative results arising from a lime phantom incorporating a titanium screw (a). (b) SEMS image arising from the green axial slice indicated in (a). (c-e) 2D imaging results delivered for the same slice by different single-shot sequences with identical FOV and resolution settings. (f, g) Images from a same acquisition involving partial sampling of the readout dimension, processed with and without POCS reconstruction. (h, i) Idem but with and without pFT$_y$ reconstruction along the xSPEN dimension. Both (g) and (i) have the same resolution as (e) but higher SNR, as evaluated by taking an average of the yellow/red squares denoting noise/signal regions shown in panel (e). FOV = 40×40 mm$^2$, thickness = 4 mm, TR = 2 s, $T_a$ = 22.02, 15.88 and 13.76 ms for (e), (f) and (h), respectively. Matrix sizes for images in (b-d) were 64×64; image sizes for the xSPEN experiments are as indicated.

Figure 3 demonstrates another aspect of how pFT procedures can improve xSPEN's sensitivity, this time focusing on trade-offs between resolution and SNR. Shown on the first row are images recorded for the same phantom and slice as introduced in Figure 2b, using the original xSPEN sequence as function of increasing matrix size. Notice how quickly this sequence trades SNR for resolution (Figs. 3a-d); this reflects the decreasing voxel sizes, but also the diffusion and relaxation penalties incurred upon seeking to increase resolution along the low-bandwidth dimension. Images reconstructed using pFT$_y$ can clearly increase SNR vis-a-vis conventionally-acquired xSPEN counterparts (Fig. 3e-h). Moreover, notice that the higher the resolution desired, the larger the SNR benefit arising from relying on a pFT.

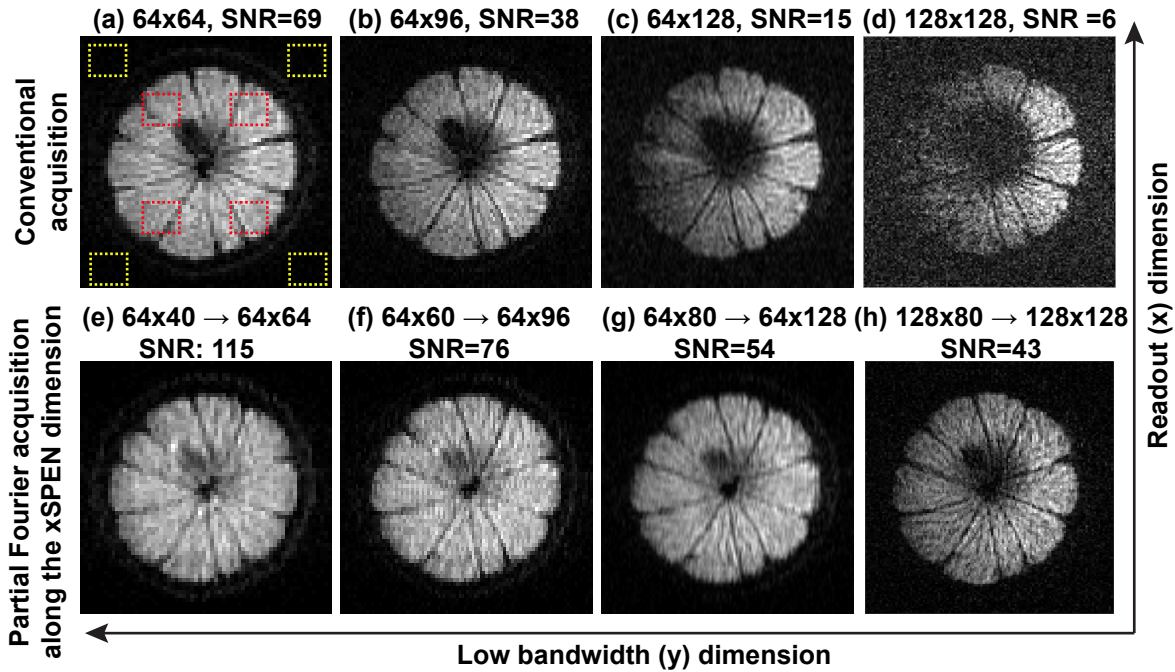

**Figure 3**. Sensitivity benefits arising from pFT along the xSPEN dimension (pFT$_y$), as judged by the SNR figures arising from the indicated yellow/red squares on the phantom introduced in Figure 2. (a-d) Images acquired with conventional xSPEN, showing how SNR is degraded as image resolution increases due to the longer $T_a$ and associated diffusion losses. (e-h) pFT$_y$ reconstructed counterparts showing how SNR gains improves with resolution. FOV = 40×40 mm$^2$, thickness = 4 mm, TR = 2 s, $T_a$ = 22.02, 33.02, 44.03, 76.8, 13.76, 20.64, 27.52 and 48 ms for (a-h) respectively, matrix sizes are as indicated.

These pFT advantages are recapitulated in Figure 4 by *in vivo* experiments on animals, which compare SEMS reference data against single-shot xSPEN images targeting a mouse head. Notice the absence of distortions in regions that typically challenge single-shot applications; e.g. near eyeballs and the ears. Notice as well the large (≥5x) SNR improvements brought about by the pFT$_y$ procedure for $p=0.625$ and the 300x300μm$^2$ in-plane resolutions that were here targeted.

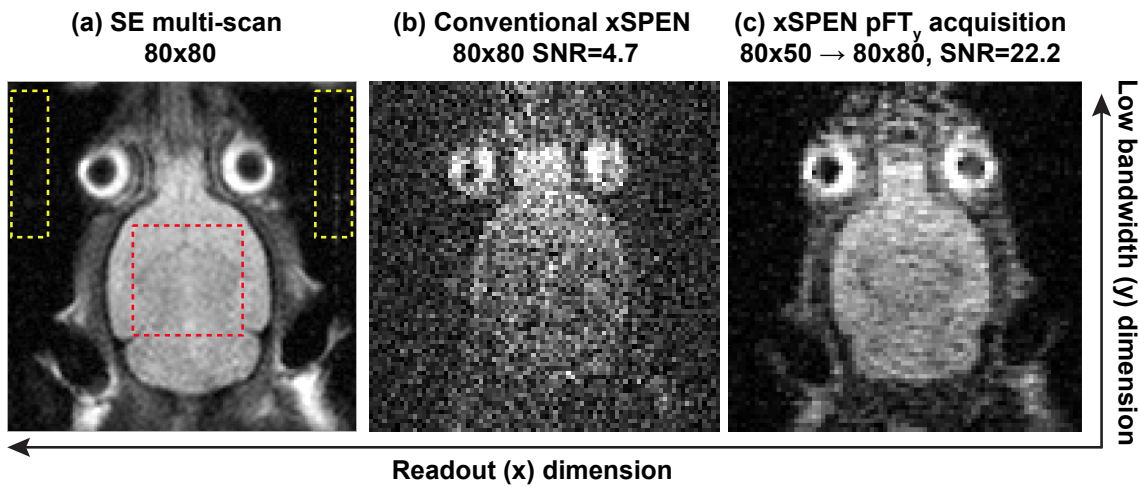

**Figure 4.** Sensitivity benefits arising from xSPEN's pFT$_y$, illustrated with *in-vivo* mice head scan. (a) Reference SEMS image acquired in 2 min 40 sec without respiration trigger, and indicating the regions used to evaluate signal (yellow) and noise (red). Images with lower and with improved SNR acquired by single-shot xSPEN MRI without (b) and with (c) pFT, so as to deliver the same resolution. FOV = 24×24 mm$^2$, slice thickness = 2.5 mm, TR = 2 s, $T_a$ = 32.6 and 20.4 ms for (b) and (c), matrix sizes as indicated.

Figure 5 illustrates a similar advantage, but for a series of scans collected at 3T on a human volunteer possessing, focusing on the frontal orbital cortex. Due to the susceptibility gradients introduced by the sinuses and eye sockets, single-shot EPI exhibits substantial distortions over various regions of the head (Fig. 5a). xSPEN yields distortion free images for this region, but the strong diffusion-driven losses arising when seeking in-plane resolutions better than 2×2 mm$^2$, render this approach of limited value (data not shown) even if restricting the FOV in order to limit the overall acquisition times. Figure 5b shows how single-shot xSPEN can successfully target these zoomed regions at thus resolution, upon introducing pFT$_y$. Although this procedure sampled only 62.5% of the readout lines, it achieves acceptable SNR and yields undistorted single-shot zoomed images free from folding and/or susceptibility artifacts.

**(a) Single-shot EPI**

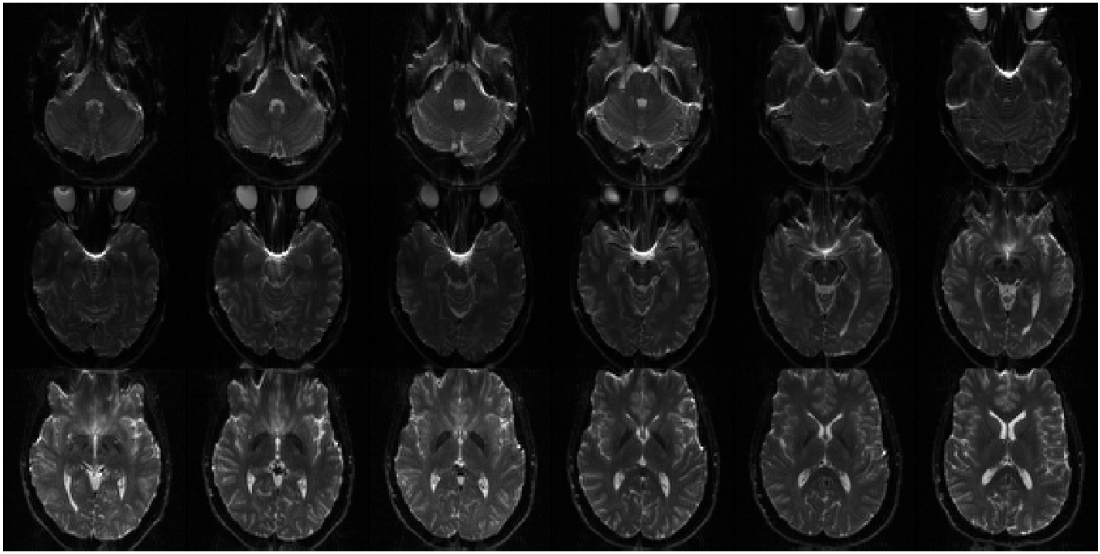

**(b) Partial Fourier reconstruction of single-shot xSPEN images with restricted FOV**

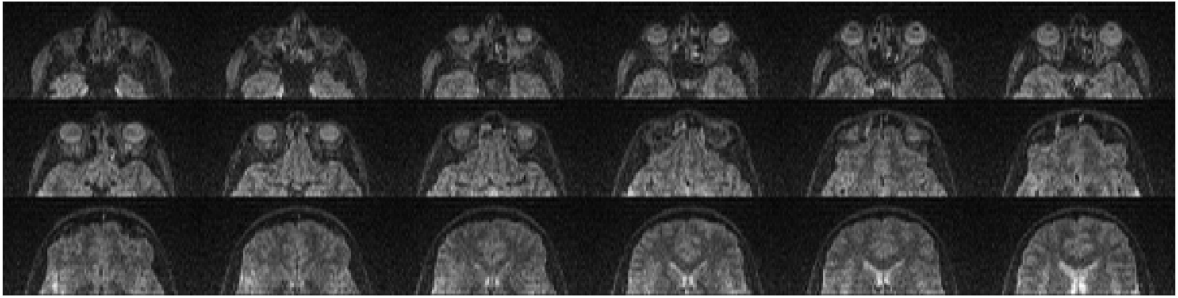

**Figure 5.** (a) Multislice single-shot EPI images (TR=2s) collected on a human volunteer at 3T; FOV = 192×192 mm$^2$, matrix size = 96×96, TE=77ms, $T_a$= 72.96 ms. (b) Multislice single-shot xSPEN images arising from the same volunteer upon performing a partial FT scan (TR=4s) along the spatiotemporal dimension; FOV = 192(RO)×96(xSPEN) mm$^2$, matrix size = 96×30 reconstructed into a 96×48 array by pFT$_y$, $T_a$ = 22.08 ms. All images possess identical 2x2 mm$^2$ in-plane resolutions.

**DISCUSSION AND CONCLUSIONS**

Single-scan xSPEN MRI suffers from SNR penalties due to its non-Fourier nature; but its main sources of signal losses are actually the diffusion and T$_2$ losses acting over its relatively long course. This is especially true when seeking a resolution improvement along the spatiotemporally encoded dimension, for which resolution is given by $\delta y = \frac{2\pi \text{FOV}_y}{T_a \cdot \gamma G_z L_z}$ (1). It follows that $\delta y$ can be improved with little or no penalty by restricting $FOV_y$; in-plane resolution can also be improved by increasing the slice thickness $L_z$, albeit at the expense of loosing out-of-plane information. Additional parameters available for increasing resolution are $G_z$, the encoding/decoding gradient that in xSPEN stays on for the course of the scan, and the acquisition time $T_a$ that, owing to xSPEN's refocusing demands, will be proportional to the each spin-packet's echo time TE. Improving resolution by increasing either of these parameters will incur in diffusion losses associated to xSPEN's constant $G_z$ gradient. Based on the Bloch-Torrey model (24,25) these losses can be approximated by an exponential attenuation varying as the square of the gradient and the cube of the free evolution time. On the basis of this, and disregarding for simplicity the effects of the refocusing pulses or of the $\pm G_y$ encoding and $\pm G_x$ readout gradients, xSPEN's diffusion-driven attenuation will be proportional to $e^{-D\gamma^2 G_z^2 (\alpha \cdot T_a)^3 / 12}$, where α is a factor depending on the voxel's position and $D$ is the diffusion coefficient.

As mentioned, pFT increases sensitivity without sacrificing image resolution despite collecting a fraction $p<1$ of the points that would normally be required. This saving can be implemented by partially sampling the readout (*x*) or the spatiotemporally encoded (*y*) dimensions. For the sequence in Fig. 1 will thus involve acquiring a *p*-fraction of the original $k_{ro}$ points, while reconstructing back the original *x*-axis resolution. Disregarding for simplicity the complications associated to ramp sampling and to the finite $\pm G_x$ gradient slew rates, pFT$_x$ will reduce the number of sampled readout points by a $p<1$, shortening accordingly the associated acquisition time $T_a$. This will lead to a reduction of the T$_2$-driven relaxation losses by $e^{-p(\alpha \cdot T_a)/T_2}$; if this is to be done without a concomitant loss in the *y*-axis resolution, however, the relation given earlier for $\delta y$ implies that $G_z$ will have to increase by a $1/p$ factor in order to keep image quality. The ensuing diffusion-related attenuation factor therefore changes from the expression given above into $e^{-p \cdot D\gamma^2 G_z^2 (\alpha \cdot T_a)^3 / 12}$; as $p<1$, this is clearly an improvement over the original attenuation factor. Compare this with the case of pFT$_y$, where the $T_a$ reduction is achieved by directly sampling fewer points along *y*-axis –i.e., by applying fewer $\pm G_x$ readout oscillations. The reduction in T$_2$-driven attenuation losses will remain as for pFT$_x$; however, the fact that the $G_z$ can now be kept at its original strength without incurring in a $\delta y$-degradation, means that the diffusion-driven attenuation factor will be reduced by $e^{-p^3 \cdot D\gamma^2 G_z^2 (\alpha \cdot T_a)^3 / 12}$. Therefore, although both

pFT$_x$ and pFT$_y$ will improve xSPEN's SNR over its original realization, pFT$_y$ is expected to a better SNR improvement due to the presence of a $p^3$- rather than a $p$-factor, in its diffusion-weighting exponent. This expectation is realized by the SNR calculations deriving from the data in Figure 2. It also explains the observations in Figure 3, whereby the higher the resolution being sought, the more there is to be gained by adopting the pFT$_y$ procedure: increases in resolution will generally be associated with the use of longer encoding and acquisition times, that will rapidly increase the diffusion-related attenuation exponent. And the the larger this exponent, the more remarkable will the effects of the $p^3$ pFT$_y$ scaling be in the final SNR.

In addition to pure SNR considerations a number of technical factors point to the convenience of choosing partial Fourier sampling along the spatiotemporal rather than the readout axis, particularly when considering xSPEN realizations in clinical settings. One of this pertains the limited *p*-reductions that can be achieved in human scanners along the readout axis, where minimum readout times are already constrained by the maximal slew rates that instrumental and physiological considerations allow one to achieve. Another limitation derives from the aforementioned need to increase the value of $G_z$ by *1/p*, upon performing pFT$_x$ without decreasing the *y*-axis image resolution. This gradient increase means that chirped pulses with larger bandwidths will be needed to cover the original $FOV_y$ and $L_z$ dimensions, resulting in a concomitant increase in the already high xSPEN SAR values.

In terms of data post-processing, however, reverse considerations apply. In particular, the fact that pFT$_x$ barely change the original simplicity of the xSPEN processing while pFT$_y$ requires both even/odd and "zigzag" phase corrections, is a factor favoring the first of these procedures. Furthermore, there is to some extent an approximation in the assumption that the even/odd and "zigzag" phase corrections can be treated independently: a more rigorous analysis of even/odd mismatch problems incorporating the zigzag effect, suggests that it is not always feasible to factor out the phase terms $e^{\pm i\beta k_x k_y^0/C}$ from the integrals introduced in Eq. (8). When this is the case (and this will naturally depend on the nature of the even/odd mismatches) artifacts can arise in images processed as described above. A general solution to this problem consists of replacing the continuous $G_z$-driven xSPEN decoding by equivalent gradient blips, acting during the ramp-times of the oscillating readout gradient train.

In summary, partial FT approaches acting along either the readout or the spatiotemporally-encoded dimensions were introduced, and shown to significantly improve the tradeoffs between resolution and SNR in single-scan xSPEN MRI. Details on how to implement these approaches were derived, and associated data processing considerations were introduced. In all cases, examples collected on preclinical and clinical scanners demonstrated the advantages of the method, while at the same time their preservation of xSPEN's unprecedented resilience to field inhomogeneities. From a practical standpoint this should unambiguously help to expand the potential applications of this emerging single-scan imaging technique. From a conceptual standpoint, the new physical insight associated to the $pFT_y$ can lead to altogether new sampling considerations of single- and multi-shot xSPEN –as will be further detailed in an upcoming publication.

**Acknowledgments.** We are grateful to Dr. Sagit Shushan (Wolfson Medical Center) and Edna Furman-Haran (Weizmann) for assistance in the human scans, and to Dr. Gilad Liberman (Weizmann) for assistance in the coding of the 3T experiments. ZZ thanks Israel's Council of Higher Education and to the Koshland Foundation for partial postdoctoral fellowships. Financial support from the Israel Science Foundation grant 795/13, the EU through ERC-2016-PoC grant # 751106, Minerva funding (#712277) from the Federal German Ministry for Education and Research, the Kimmel Institute for Magnetic Resonance and the generosity of the Perlman Family Foundation, are also acknowledged.

**Figure Captions**